\documentclass[journal]{IEEEtran}
\usepackage{cite}
\usepackage{amsmath,amssymb,amsfonts}
\usepackage{algorithm}
\usepackage{algorithmic}
\usepackage{makecell}
\usepackage{threeparttable}
\usepackage{graphicx}
\usepackage{textcomp}
\usepackage{xcolor}
\usepackage{multirow}
\usepackage{dsfont}

\ifCLASSINFOpdf

\else

\fi

\begin{document}

\title{All-Around Real Label Supervision: \\Cyclic Prototype Consistency Learning for Semi-supervised Medical Image Segmentation}

\author{Zhe Xu, Yixin Wang, Donghuan Lu, Lequan Yu, \IEEEmembership{Member, IEEE}, Jiangpeng Yan, Jie Luo, Kai Ma, Yefeng Zheng, \IEEEmembership{Fellow, IEEE}, and Raymond Kai-yu Tong, \IEEEmembership{Senior Member, IEEE}

\thanks{This research was done with Tencent Jarvis Lab and partly supported by General Research Fund from Research Grant Council of Hong Kong (No. 14205419), Key-Area Research and Development Program of Guangdong Province, China (No. 2018B010111001), and the Scientific and Technical Innovation 2030-``New Generation Artificial Intelligence” Project (No. 2020AAA0104100). \textsl{Corresponding authors: Raymond Kai-yu Tong (e-mail: kytong@cuhk.edu.hk) and Donghuan Lu (e-mail: caleblu@tencent.com)}.}
\thanks{Z. Xu and R. Tong are with Department of Biomedical Engineering, The Chinese University of Hong Kong, Shatin, NT, Hong Kong, China. (e-mail: jackxz@link.cuhk.edu.hk; kytong@cuhk.edu.hk).}
\thanks{D. Lu, K. Ma and Y. Zheng are with Tencent Jarvis Lab, Shenzhen, China. Y. Wang is with Institute of Computing Technology, Chinese Academy of Sciences, Beijing, China. L. Yu is with Department of Statistics and Actuarial Science, The University of Hong Kong, Hong Kong, China. J. Yan is with Department of Automation, Tsinghua University, Beijing, China. J. Luo is with Brigham and Women's Hospital, Harvard Medical School, Boston, MA, USA.}}

\markboth{}
{Xu \MakeLowercase{\textit{et al.}}: Cyclic Prototype Consistency Learning for Semi-supervised Medical Image Segmentation}

\maketitle

\begin{abstract}
Semi-supervised learning has substantially advanced medical image segmentation since it alleviates the heavy burden of acquiring the costly expert-examined annotations. Especially, the consistency-based approaches have attracted more attention for their superior performance, wherein the real labels are only utilized to supervise their paired images via supervised loss while the unlabeled images are exploited by enforcing the perturbation-based \textit{``unsupervised"} consistency without explicit guidance from those real labels. However, intuitively, the expert-examined real labels contain more reliable supervision signals. Observing this, we ask an unexplored but interesting question: can we exploit the unlabeled data via explicit real label supervision for semi-supervised training? To this end, we discard the previous perturbation-based consistency but absorb the essence of non-parametric prototype learning. Based on the prototypical networks, we then propose a novel cyclic prototype consistency learning (CPCL) framework, which is constructed by a labeled-to-unlabeled (L2U) prototypical forward process and an unlabeled-to-labeled (U2L) backward process. Such two processes synergistically enhance the segmentation network by encouraging more discriminative and compact features. In this way, our framework turns previous \textit{``unsupervised"} consistency into new \textit{``supervised"} consistency, obtaining the \textit{``all-around real label supervision"} property of our method. Extensive experiments on brain tumor segmentation from MRI and kidney segmentation from CT images show that our CPCL can effectively exploit the unlabeled data and outperform other state-of-the-art semi-supervised medical image segmentation methods.
\end{abstract}


\section{Introduction}
\label{sec:introduction}
Segmenting anatomical organs or tumors from medical images plays an important role in many image-guided therapies. Recently, deep learning (DL)-based approaches have achieved satisfactory performance in various medical image analysis tasks \cite{li2018h,dou20163d,chen2020realistic,xu2020adversarial}. However, the success of DL-based segmentation approaches usually relies on a large amount of labeled data, which is particularly difficult and costly to obtain in the medical imaging domain where only experts can provide reliable and accurate annotations and the images are often in 3D volumes. The heavy burden of acquiring the costly expert-examined labels motivates many annotation-efficient researches such as semi-supervised learning (SSL) \cite{tarvainen2017mean,zhang2017DAN,sedai2017GANsemi,vu2019EM,verma2019ICT,luo2020DTC}, weakly supervised learning \cite{xu2021noisy,yang2020weakly}, and self-supervised learning \cite{zhuang2019self,chen2019self}. In this work, we focus on semi-supervised segmentation since it is clinically practical to obtain a small set of expert-examined labeled data and a large quantity of unlabeled images.

\begin{figure}[t]
\centerline{\includegraphics[width=\columnwidth]{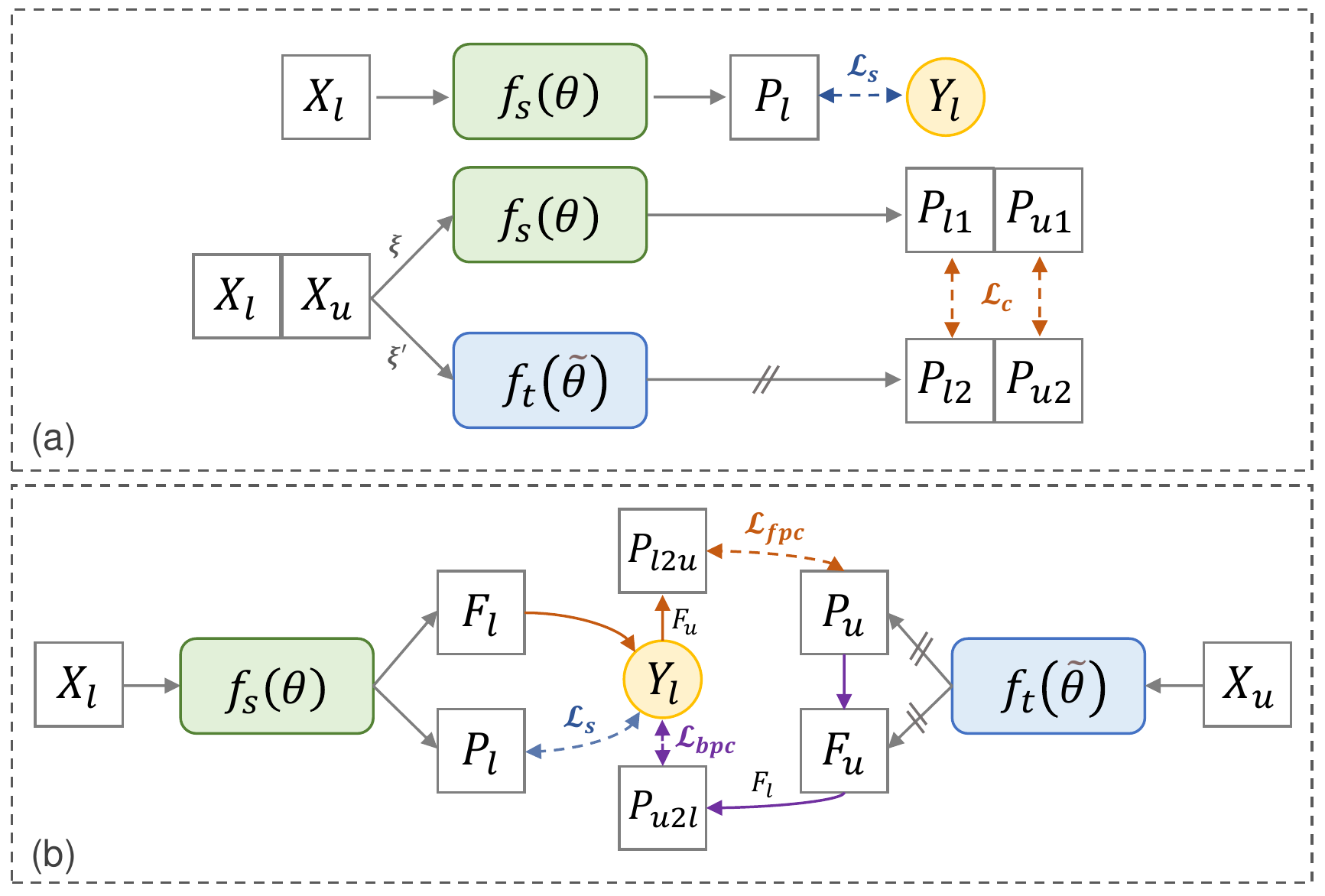}}
\caption{Comparison between (a) the typical perturbed consistency-based self-ensembling SSL paradigm and (b) our proposed cyclic prototype consistency learning (CPCL) paradigm. Both paradigms adopt a student model $f_{s}(\theta)$ and a self-ensembling teacher model $f_{t}(\tilde{\theta})$. $X_{l}$ and $X_{u}$ denote the labeled images and unlabeled images, respectively, along with corresponding real label $Y_{l}$ for $X_{l}$. $\xi$ and $\xi^{\prime}$ in (a) denote different perturbations (e.g., random Gaussian noises). $P$ represents the segmentation prediction and $F$ in (b) denotes the features extracted from the encoder. Note that the features $F$ on the arrows indicate that they are the to-be-segmented features. ‘$//$’ means stop-gradient. Obviously, the main difference is that the typical paradigm utilizes the costly pixel-wise real labels $Y_{l}$ only for supervising the corresponding labeled data, i.e., $\mathcal{L}_{\boldsymbol{s}}$, while the real label supervision can circulate throughout our CPCL with the help of non-parametric cyclic prototype learning, that is, the unlabeled data can be exploited via explicit real label supervision. Detailed architecture of CPCL is illustrated in Sec. \ref{sec:method}.}
\label{fig_intro_comparison}
\end{figure}

Recent impressive progress in semi-supervised medical image segmentation has featured the smoothness assumption \cite{luo2018smooth} based consistency learning \cite{li2020transformation,yu2019uncertainty,zhang2021DTML}. Besides the supervised loss for labeled data, this paradigm leverages unlabeled data by enforcing \textit{``unsupervised"} perturbation-based consistency between predictions of the self-ensembling teacher model and the student model, as shown in Fig. \ref{fig_intro_comparison} (a). Based on it, some improved works, e.g., using more suitable types or strengths of the perturbation \cite{li2020transformation,verma2019ICT,french2019semi,xie2019UDA} and adopting uncertainty to select more reliable voxel-wise consistency targets \cite{yu2019uncertainty,wang2020double}, are proposed. Intuitively, the quality of the unsupervised perturbed consistency target determines whether such paradigm can successfully exploit the intrinsic information in unlabeled images. As such, this paradigm may have the following limitations that can lead to sub-optimal results: (i) difficulty to define the most suitable types or strengths of the perturbation for different tasks that will greatly affect the performance; (ii) limited reliability during training: since the real labels are only provided to their paired labeled images, the unlabeled data cannot benefit from explicit expert-examined supervision. However, distinct from these examined labels, the reliability of the unsupervised consistency targets is hard to guarantee.

\begin{figure*}[t]
\centerline{\includegraphics[width=2\columnwidth]{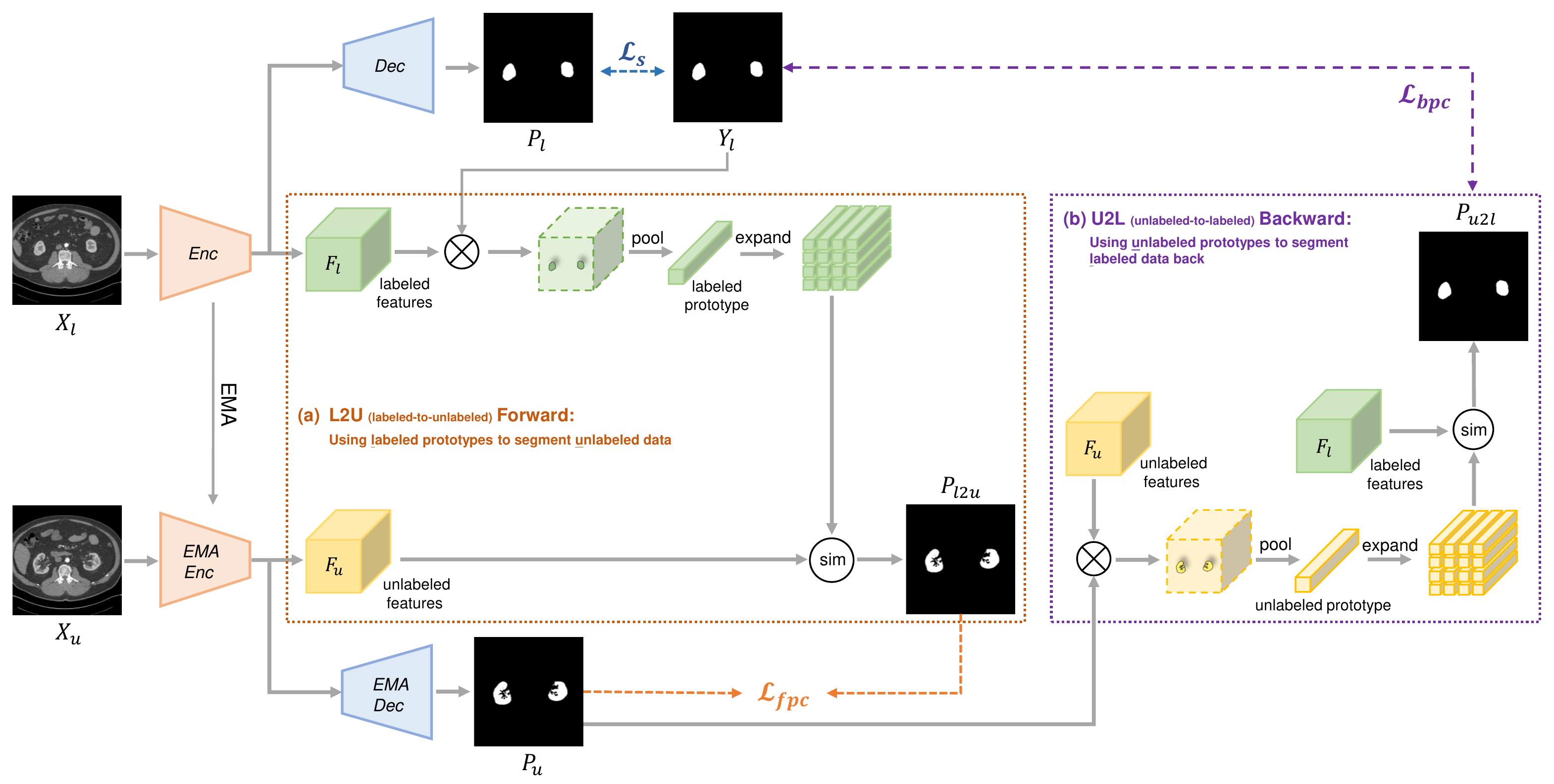}}
\caption{Illustration of the proposed cyclic prototype consistency learning (CPCL) framework for semi-supervised medical image segmentation (using kidney segmentation as an example). \textsl{Enc} and \textsl{Dec} represent encoder and decoder of the student model, while \textsl{EMA Enc} and \textsl{EMA Dec} denote the self-ensembling teacher model updated as the exponential moving average (EMA) of student weights. The feature maps $F_{l}$ and $F_{u}$ are extracted from the last convolution layer of the encoder and upsampled to match the size of the segmentation mask by trilinear interpolation operation. $\otimes$ denotes the element-wise multiplication operation. ``sim" denotes the cosine similarity calculation between each prototype and to-be-segmented features at each spatial location, followed by a softmax operation to produce the probability map. Apart from the supervised loss ($\mathcal{L}_{s}$) on labeled data, the framework is also supervised by a real label-centric cyclic consistency learning mechanism, which is consisted of (a) labeled-to-unlabeled prototypical forward process (contributing to $\mathcal{L}_{fpc}$), and (b) unlabeled-to-labeled prototypical backward process (contributing to $\mathcal{L}_{bpc}$).}
\label{fig_framework}
\end{figure*}

Observing the above limitations, it is natural to ask the following question: \textit{can we exploit the unlabeled data via explicit real label supervision for semi-supervised training?} To this end, first, we discard the previous perturbation-based consistency but absorb the essence of non-parametric prototype learning \cite{snell2017prototypical,dong2018few} commonly used in few-shot learning. Based on the prototypical networks, we then propose a novel real label-centric cyclic prototype consistency learning (CPCL) framework for semi-supervised segmentation. The paradigm is shown in Fig. \ref{fig_intro_comparison} (b) and detailed architecture is shown in Fig. \ref{fig_framework}. Specifically, on top of the self-ensembling strategy, the proposed CPCL consists of an L2U forward process, i.e., using \underline{l}abeled prototypes to segment \underline{u}nlabeled data (Fig. \ref{fig_framework} (a)), and a U2L backward process, i.e., using \underline{u}nlabeled prototypes to segment \underline{l}abeled data back (Fig. \ref{fig_framework} (b)). When the unlabeled data are fed into the self-ensembling model, the L2U forward consistency targets are driven by the real label prototypes and the extracted features from unlabeled data. Synergistically, the U2L backward process utilizes the unlabeled prototypes, which are driven by the learned unlabeled features and the unlabeled prediction from the self-ensembling model, to perform prototypical segmentation for labeled data back, so that the backward consistency target can be directly guided by the reliable real label. Such a cyclic scheme enhances the segmentation network by encouraging the network learning more discriminative and compact representation from labeled and unlabeled data, and turns previous \textit{``unsupervised"} consistency into new \textit{``supervised"} consistency, obtaining the \textit{``all-around real label supervision"} property of our method. 

Overall, the main contributions of this work are as follows:
\begin{itemize}
    \item We present a new perspective for semi-supervised medical image segmentation, that is, exploiting the unlabeled data via explicit real label supervision.
    \item Based on the above perspective, we propose a real label-centric cyclic prototype consistency learning (CPCL) framework for semi-supervised training, which turns previous \textit{``unsupervised"} consistency to a new \textit{``supervised"} consistency with the help of prototypical networks.
    \item We have conducted extensive experiments on brain tumor segmentation from magnetic resonance imaging (MRI) and kidney segmentation from computed tomography (CT) images. The comparison and ablation studies demonstrate the superiority of our CPCL over other state-of-the-art semi-supervised methods and the effectiveness of each component.
\end{itemize}

\section{Related Work}
\subsection{Semi-supervised Medical Image Segmentation}
To reduce the annotation efforts, semi-supervised medical image segmentation has been studied for a long period. In most early semi-supervised works, they perform segmentation with the help of hand-crafted features. For example, \textit{You et al.} \cite{you2011segmentation} presented a prior-based method to improve retinal vessel segmentation from fundus images, and \textit{Portela et al.} \cite{portela2014semi} proposed a clustering-based Gaussian mixture model to segment brain MR images. However, these approaches often suffer from unsatisfactory performance due to the limited representation capacity of the hand-crafted features. 

With a stronger ability to automatically learn high-level representation, deep learning has greatly advanced semi-supervised medical image segmentation. \textit{Bai et al.} \cite{bai2017semi} proposed a self-training approach for cardiac segmentation from MRI, which iteratively updates the network parameters and pseudo labels for unlabeled data. Then, multi-view co-training is explored for semi-supervised liver segmentation \cite{zhou2019semicotrainig} and breast cancer analysis \cite{xia2020cotrainig}. Adversarial learning also becomes a popular solution for semi-supervised segmentation \cite{zhang2017DAN,nie2018GANasdnet,sedai2017GANsemi}. As a typical example, \textit{Zhang et al.} \cite{zhang2017DAN} proposed a deep adversarial network (DAN) for biomedical image segmentation by encouraging the predicted segmentation of unlabeled data to be similar to that of labeled data. 

More recently, consistency learning, which leverages unlabeled data by enforcing unsupervised perturbation-based consistency, achieves impressive performance in semi-supervised learning. For example, the $\Pi$ model \cite{samuli2017temporal} performs multiple forward predictions under different perturbations and encourages consistency of the network outputs. Then, the temporal ensembling strategy \cite{samuli2017temporal} improves the $\Pi$ model by using \textit{exponential moving average} (EMA) predictions for unlabeled data as the consistency targets. However, maintaining the EMA predictions becomes a heavy burden during training. To cope with it, the mean teacher (MT) framework \cite{tarvainen2017mean,cui2019semi} proposed to use a teacher model with the EMA weights of the student model. Inspired by the MT model, some recent improved works were further proposed. \textit{Yu et al.} \cite{yu2019uncertainty} extended the MT model by introducing uncertainty map to select reliable voxels as the consistency targets. \textit{Luo et al.} \cite{luo2021efficient} extended the uncertainty rectified consistency with pyramid multi-scale strategy. \textit{Li et al.} \cite{li2020transformation} introduced a transformation-consistency to further improve the semi-supervised skin lesion segmentation. \textit{Luo et al.} \cite{luo2020DTC} and \textit{Zhang et al.} \cite{zhang2021DTML} combined the regression task with the segmentation task to form a dual-task consistency and thereby explicitly impose geometric constraints. \textit{You et al.} \cite{you2021simcvd} introduced contrastive loss as the auxiliary supervision for unlabeled data. Based on contrastive loss, \textit{Lai et al.} \cite{lai2021semiCVPR21} constructed a context-aware consistency to make representations more robust to the environment. Generally, to exploit the unlabeled data, recent consistency-based works are similarly dedicated to investigating better perturbation (e.g., noises and transformations) for unsupervised consistency learning or introducing other parallel tasks to provide auxiliary guidance. However, the existing works utilized the expert-examined labels only for the labeled data training, while the unlabeled data training regrettably cannot obtain any explicit guidance from those costly labels. As another alternative, we present a new perspective that exploit the unlabeled data via the trustworthy real label, i.e., turning existing \textit{``unsupervised"} consistency to a new \textit{``supervised"} consistency.

\begin{figure}[t]
\centerline{\includegraphics[width=0.9\columnwidth]{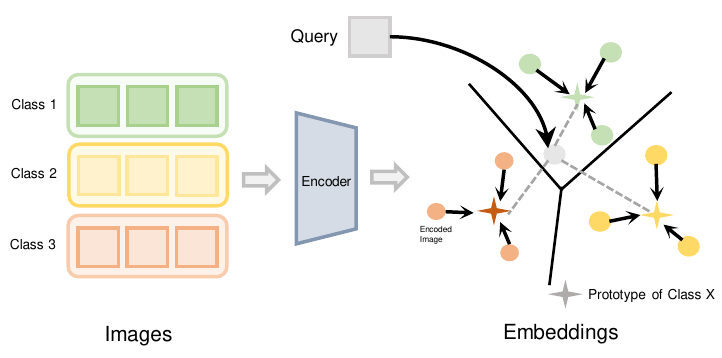}}
\caption{Concise diagram of the prototypical networks in image classification \cite{snell2017prototypical}. The prototypical networks aim to obtain well-separated per-class prototypes (stars) in the feature space so that we can classify each query image by comparing the distances (dotted line) between its embedded features and class prototypes.}
\label{fig_pronet}
\end{figure}

\subsection{Prototype Learning}
Apparently, the above perspective poses a major challenge: the expert-examined real labels and the unlabeled data are unpaired. In our work, we mainly draw on the spirit of non-parametric prototype learning in few-shot segmentation to solve this problem. Few-shot segmentation aims to segment targets with only a few labeled samples, wherein the prototype-based approaches perform pixel-wise matching on query images with holistic prototypes of different classes extracted from support set \cite{snell2017prototypical,zhang2020late,yang2020mixprototype,wang2019panet,li2021adaptive,dong2018few}. Specifically, the prototypical networks \cite{snell2017prototypical} were firstly proposed to learn a metric space and perform image classification via computing distances to the class-related prototypes. Concisely, such strategy relies on the idea that there exists an embedding space in which relevant features cluster around a representative prototype for each class, where the prototypical networks aim to learn per-class prototypes on top of sample averaging in the feature space. As shown in Fig. \ref{fig_pronet}, each prototype is the mean vector of the embedding features belonging to its class and can be regarded as the representative of this class. Thereby, the classification can be performed by computing distances between the prototype of each class and the embedding features of the query images. Intuitively, such design encourages more discriminative and compact features towards more appealing classification performance. Following this spirit, \textit{Dong et al.} \cite{dong2018few} further adapted the prototypical networks to tackle the few-shot pixel-wise classification task, i.e., image segmentation. SG-One \cite{zhang2020late} designed a masked average pooling strategy to obtain the squeezed representation of support images and then applied the cosine similarity to perform pixel-wise matching between pixels in query images and the prototypes. PANet \cite{wang2019panet} further introduced a backward prototype alignment between support and query branches to regularize the prototype learning. ASGNet \cite{li2021adaptive} introduced the superpixel technique to form an adaptive prototype allocation mechanism. FWB \cite{nguyen2019FWB} improved the quality of support prototypes by considering the support feature differences between foreground and background. To tackle the challenge caused by unpaired supervision, we bridge semi-supervised and few-shot segmentation tasks with prototype learning, where the labeled set can be regarded as a support set while the unlabeled images as query images, respectively.

\section{Methodology}
\label{sec:method}
Fig. \ref{fig_framework} depicts the proposed cyclic prototype consistency learning (CPCL) framework for semi-supervised medical image segmentation. In the existing consistency-based SSL methods, the expert-examined real labels $Y_{l}$ are only exploited for their corresponding images by means of the supervised loss $\mathcal{L}_{\boldsymbol{s}}$, as shown in Fig. \ref{fig_intro_comparison}(a). Here, we give a new perspective for semi-supervised segmentation. We aim to make full use of those costly pixel-wise real labels, that is, both labeled and unlabeled data can obtain effective guidance from the expert-examined real labels. To this end, we absorb the essence of prototype learning commonly used in few-shot segmentation and form a real label-centric cyclic consistency learning framework (conceptual paradigm is shown in Fig. \ref{fig_intro_comparison}(b)) by constructing the labeled-to-unlabeled (L2U) prototypical forward process (Fig. \ref{fig_framework}(a)) and the unlabeled-to-labeled (U2L) prototypical backward process (Fig. \ref{fig_framework}(b)). Noteworthily, the proposed CPCL does not employ any perturbation. The detailed description of the proposed framework is elaborated in the following sections.

\subsection{Framework Overview}
\label{sec:overview}
To ease the description of the methodology, we first formulate the semi-supervised segmentation problem. In this task, the training set consists of $N$ samples in total, while only $M$ samples have the expert-examined real labels and the remaining $N-M$ samples only include the images, e.g., MRI. We denote the labeled set as $\mathcal{S}_{L}=\left\{\left(X_{l(i)}, Y_{l(i)}\right)\right\}_{i=1}^{M}$ and the unlabeled set as $\mathcal{S}_{U}=\left\{X_{u(i)}\right\}_{i=M+1}^{N}$, where $X_{l(i)}, X_{u(i)} \in \mathbb{R}^{H \times W \times D}$ represent the input 3D volumes of height $H$, width $W$, depth $D$ and  $ Y_{l(i)} \in\{0,1\}^{H \times W \times D}$ 
is the ground-truth segmentation label of $X_{l(i)}$. Although the core spirit of our CPCL is substantially different from previous perturbation-based approaches, the proposed method can still be regarded as another type of consistency regularization, i.e., we utilize the real label supervision signals to regularize the learning from the unlabeled set. Therefore, this task can be formulated as training the network by optimizing the following objective:
\begin{equation}
\label{eq:overview}
\min _{\theta}  \mathcal{L}_{s}\left(\theta,\mathcal{S}_{L}\right)+\lambda \mathcal{L}_{c}(\theta, \tilde{\theta}, \mathcal{S}_{L}, \mathcal{S}_{U}),
\end{equation}
where $\mathcal{L}_{s}$ is the supervised loss for labeled data; $\mathcal{L}_{c}$ represents the general consistency loss, e.g., the proposed cyclic prototype consistency in our work or the perturbation-based prediction consistency in previous works \cite{cui2019semi,tarvainen2017mean}; $\theta$ and $\tilde{\theta}$ denote the weights of the student model and the teacher model, respectively; $\lambda$ is a ramp-up weight commonly scheduled by the time-dependent Gaussian function $\lambda(t)=w_{max} \cdot e^{\left(-5\left(1-\frac{t}{t_{m a x}}\right)^{2}\right)}$ to make the tradeoff between $\mathcal{L}_{s}$ and $\mathcal{L}_{c}$ \cite{cui2019semi}, where $w_{max}$ is the final consistency weight and $t_{\max}$ is the maximum number of training steps. Such design for $\lambda$ can avoid the domination by meaningless consistency targets at the beginning of network training.  

Besides, recent progress on semi-supervised segmentation indicates that constructing a self-ensembling teacher model from the student model at different training steps can enhance the reliability of the teacher model's prediction \cite{cui2019semi}. Thus, following this spirit, we update the teacher model's weights $\tilde{\theta}_{t}$ at the training step $t$ by means of the \textit{exponential moving average} (EMA) approach, which can be formulated as:
\begin{equation}
\label{eq:EMA}
\tilde{\theta}_{t}=\alpha \tilde{\theta}_{t-1}+(1-\alpha) \theta_{t},
\end{equation}
where $\alpha$ is the EMA decay rate and set to 0.99 as recommended by \cite{tarvainen2017mean,xu2021noisy}.

\subsection{Cyclic Prototype Consistency Learning}
As mentioned in Sec. \ref{sec:introduction}, the previous unsupervised targets easily suffer from limited reliability. Without expert-examined supervision, it is difficult to obtain precise segmentation masks from unlabeled images. Instead of constraining unsupervised consistency like in previous approaches, we propose a cyclic prototype consistency learning schema on top of the prototypical networks \cite{snell2017prototypical} to encourage network to learn robust well-representative and well-separated prototype representation for the segmentation targets and concurrently realizing contextual information incorporation between labeled and unlabeled data. Specifically, since the voxel embeddings belonging to the same segmentation targets should be similar, we adopt the masked average pooling operation with the late prototype generation strategy \cite{zhang2020late} which masks the intermediate foreground/background features extracted from the encoder with the segmentation masks. These prototypes with abundant contextual information from labeled data (or unlabeled data) can be then explored to measure the similarity to segment the unlabeled data (or labeled data) via non-parametric metric learning. Such strategy is integrated into our cyclic framework including two tailored processes, i.e., L2U forward consistency and U2L backward consistency, as elaborated successively.  

\subsubsection{L2U Forward Consistency}
The purpose of L2U forward process (i.e., using labeled prototypes to segment unlabeled data) is to utilize the generated prototype to transfer the real label supervision signals from labeled data to unlabeled data, as shown in Fig. \ref{fig_framework} (a). Note that since most previous efforts are evaluated in binary segmentation tasks, we will consistently focus on binary segmentation in this work, i.e., one foreground target and one background target are considered. However, the proposed framework can be easily adapted to multi-class segmentation, i.e., generating multiple foreground prototypes for different segmentation targets. Specifically, let $F_{l(k)}$ be the feature map extracted by the encoder for the labeled image $X_{l(k)}$, where $k=1, ..., K$ indexes the images in a mini-batch during training. The corresponding real label of $X_{l(k)}$ is $Y_{l(k)}$. The prototype $p_{l\mathrm{(fg)}}$ of the foreground segmentation target $\mathcal{C_\mathrm{fg}}$ is generated via masked average pooling \cite{zhang2020late}:
\begin{equation}
\label{eq:plfg}
p_{l\mathrm{(fg)}}=\frac{1}{K} \sum_{k} \frac{\sum_{x, y, z} F_{l(k)}^{(x, y, z)} \mathds{1}\left[Y_{l(k)}^{(x, y, z)}\in\mathcal{C_\mathrm{fg}}\right]}{\sum_{x, y, z} \mathds{1}\left[Y_{l(k)}^{(x, y, z)}\in\mathcal{C_\mathrm{fg}}\right]},
\end{equation}
where the feature map $F_{l(k)}$ is upsampled to match the size of the segmentation mask by trilinear interpolation; $(x, y, z)$ denotes the spatial location for each voxel; and $\mathds{1}(\cdot)$ represents the indicator function that returns 1 when the condition is true or 0 otherwise. Note that to simplify the illustration, we omit the prototype of the background in Fig. \ref{fig_framework}. Accordingly, the prototype of background can also be generated by:
\begin{equation}
\label{eq:plbg}
p_{l\mathrm{(bg)}}=\frac{1}{K} \sum_{k} \frac{\sum_{x, y, z} F_{l(k)}^{(x, y, z)} \mathds{1}\left[Y_{l(k)}^{(x, y, z)}\notin \mathcal{C_\mathrm{fg}}\right]}{\sum_{x, y, z} \mathds{1}\left[Y_{l(k)}^{(x, y, z)}\notin \mathcal{C_\mathrm{fg}}\right]}.
\end{equation}

To learn the optimal prototypes collaboratively with the segmentation network, we adopt a non-parametric metric learning mechanism. Noteworthily, this mechanism introduces no extra learnable parameters that may lead to over-fitting. Concretely, we introduce a distance function $d(\cdot)$ to measure the similarity between the unlabeled feature map (denoted as $F_{u}$) extracted from the self-ensembling teacher model and the labeled prototypes $p_{l\mathrm{(fg)}}$ or $p_{l\mathrm{(bg)}}$. Then, the softmax function over the similarities is applied to produce the probability map $P_{l2u}$ over the classes. Let $\mathcal{P}_{l}=\left\{p_{l_\mathrm{(fg)}}\right\} \cup\left\{p_{l_\mathrm{(bg)}}\right\}$, for each $p_{l(j \in\{bg,fg\} )} \in \mathcal{P}_{l}$, we have:
\begin{equation}
\label{eq:pl2u}
P_{l2u(j)}^{(x, y, z)}=\frac{\exp \left(-\alpha d\left(F_{u}^{(x, y, z)}, p_{l(j)}\right)\right)}{\sum_{p_{l(j)} \in \mathcal{P}_{l}} \exp \left(-\alpha d\left(F_{u}^{(x, y, z)}, p_{l(j)}\right)\right)},
\end{equation}
where $d(\cdot)$ adopts the cosine distance, and multiplier $\alpha$ is a scaling factor set as 20 recommended by \cite{wang2019panet}. Note that the unlabeled feature map $F_{u}$ also undergoes the trilinear-interpolation-based upsampling process. After obtaining labeled-to-unlabeled prototypical prediction $P_{l2u}$ via metric learning, we propagate such real label signal to the unlabeled data via the forward prototype consistency loss $\mathcal{L}_{fpc}$, calculated by:
\begin{equation}
\mathcal{L}_{fpc}=\mathcal{L}_{mse}(P_{l2u}, P_{u}),
\end{equation}
where $P_{u}$ is the predicted probability of the unlabeled data from the entire self-ensembling teacher model and $\mathcal{L}_{mse}$ denotes the loss of mean squared error (MSE). In this way, the real label driven prototypical segmentation $P_{l2u}$ can serve as effective guidance for the unlabeled data to discover the desired target regions, and thereby provide more meaningful knowledge for the network training, as experimentally demonstrated in Sec. \ref{sec:ablation}. 

\subsubsection{U2L Backward Consistency}
Synergistically, another backward process is expected to discover the most representative prototype of the same target regions from unlabeled data, where more strict and direct guidance will be imposed by the labeled data via U2L backward consistency, i.e., using unlabeled prototypes to segment labeled data, as shown in Fig. \ref{fig_framework} (b). To well-roundly enhance $P_{u}$, $F_{u}$ and $F_{l}$, we use the unlabeled prediction ($P_{u}$) and the extracted feature maps ($F_{l}$ and $F_{u}$) from the above L2U forward process to deduce the prototypical predictions $P_{u2l}$ back for labeled data. By such, the real label $Y_{l}$ can be directly employed again to guide the learning from unlabeled data (i.e., $\mathcal{L}_{bpc}$ in Fig. \ref{fig_framework}), obtaining the \textit{``all-around real label supervision"} property of our method. Specifically, following Eqns. (\ref{eq:plfg}), (\ref{eq:plbg}) and (\ref{eq:pl2u}), we can reversely obtain the unlabeled-to-labeled prototypical prediction $P_{u2l}$ via non-parametric metric learning as well. First, the unlabeled binary prediction mask is given by
\begin{equation}
\hat{Y}_{u}=\underset{j\in\{bg,fg\}}{\arg \max }(P_{u(j)}).
\end{equation}

Then, the foreground and background holistic prototypes for unlabeled data can be obtained by masked average pooling:
\begin{equation}\left\{\begin{array}{l}
p_{u\mathrm{(fg)}}=\frac{1}{K} \sum_{k} \frac{\sum_{x, y, z} F_{u(k)}^{(x, y, z)} \mathds{1}\left[\hat{Y}_{u(k)}^{(x, y, z)}\in\mathcal{C_\mathrm{fg}}\right]}{\sum_{x, y, z} \mathds{1}\left[\hat{Y}_{u(k)}^{(x, y, z)}\in\mathcal{C_\mathrm{fg}}\right]}; \\
p_{u\mathrm{(bg)}}=\frac{1}{K} \sum_{k} \frac{\sum_{x, y, z} F_{u(k)}^{(x, y, z)} \mathds{1}\left[\hat{Y}_{u(k)}^{(x, y, z)}\notin \mathcal{C_\mathrm{fg}}\right]}{\sum_{x, y, z} \mathds{1}\left[\hat{Y}_{u(k)}^{(x, y, z)}\notin \mathcal{C_\mathrm{fg}}\right]}.
\end{array}\right.\end{equation}

After obtaining the unlabeled prototypes $\mathcal{P}_{u}=\left\{p_{u_\mathrm{(fg)}}\right\} \cup\left\{p_{u_\mathrm{(bg)}}\right\}$, we can further calculate the cosine distances and produce the U2L probability map $P_{u2l}$ by:
\begin{equation}
\label{eq:pu2l}
P_{u2l(j)}^{(x, y, z)}=\frac{\exp \left(-\alpha d\left(F_{l}^{(x, y, z)}, p_{u(j)}\right)\right)}{\sum_{p_{u(j)} \in \mathcal{P}_{u}} \exp \left(-\alpha d\left(F_{l}^{(x, y, z)}, p_{u(j)}\right)\right)}.
\end{equation}

Finally, the backward prototype consistency loss $\mathcal{L}_{bpc}$ is calculated by: 
\begin{equation}
\mathcal{L}_{bpc}=\mathcal{L}_{ce}(Y_{l}, P_{u2l}),
\end{equation}
where $\mathcal{L}_{ce}$ denotes the commonly used cross-entropy (CE) loss. Intuitively, if the self-ensembling teacher model predicts a good segmentation mask for the unlabeled data along with well-discriminative features $F_{u}$, the unlabeled prototypes are able to distinguish the labeled features $F_{l}$ voxel-by-voxel and thereby segment the labeled images well. Thus, the backward process further regularizes the quality of $P_{u}$ in turn to enhance the forward process. Both processes highly collaborate with each other in a cyclic scheme to turn the previous \textit{``unsupervised"} consistency to new \textit{``supervised"} consistency, and such a cycle forces the model to enhance the quality of both embedding spaces and segmentation predictions, as experimentally demonstrated in Sec. \ref{sec:ablation}.

\subsection{Semi-supervised Training}
\label{sec:loss}
The total objective function to train our cyclic prototype consistency learning framework is a weighted combination of the supervised loss $\mathcal{L}_{s}$ on labeled data only, and the forward and backward prototype consistency loss $\mathcal{L}_{fpc}$ and $\mathcal{L}_{bpc}$ driven by both labeled data and unlabeled data synergistically. Formally, the total loss is calculated as follows:
\begin{equation}
\label{eq:total_loss}
\mathcal{L}=\mathcal{L}_{s}+\lambda \mathcal{L}_{c}, \quad \text { with } \quad \mathcal{L}_{c}=\mathcal{L}_{fpc}+\beta \mathcal{L}_{bpc}
\end{equation}
where $\lambda$ is the time-dependent ramp-up weight described in Sec. \ref{sec:overview}; $\beta$ is a hyper-parameter to balance $\mathcal{L}_{fpc}$ and $\mathcal{L}_{bpc}$, which is generally set as 10 and the effect of this hyper-parameter is also studied in Sec. \ref{sec:beta}. Note that the supervised loss $\mathcal{L}_{s}$ is a combination of cross-entropy loss and Dice loss:
\begin{equation}
\label{eq:Ls}
\mathcal{L}_{s}=0.5*\mathcal{L}_{ce}(Y_{l}, P_{l})+0.5*\mathcal{L}_{Dice}(Y_{l}, P_{l}),
\end{equation}
since we found that such a combination can provide better performance under most supervised-only settings in our exploratory study. 

\section{Experiments}
We evaluate our proposed semi-supervised segmentation method on both whole brain tumor segmentation from T2 fluid attenuated inversion recovery (T2-FLAIR) MRI and kidney segmentation from arterial phase abdominal CT scans, with extensive ablation analysis and comparison study with state-of-the-art semi-supervised methods.
\subsection{Datasets and Experimental Setup}
\subsubsection{Brain Tumor Segmentation Dataset} 
The experiment of whole brain tumor segmentation is performed using the T2-FLAIR MRI data from the BraTS 2019 challenge \cite{brats19}. The entire dataset contains multi-institutional preoperative MRI of 335 glioma patients, including 259 high-grade glioma (HGG) patients and 76 low-grade glioma (LGG) patients, where each patient has four modalities of MRI scans including T1, T1Gd, T2 and T2-FLAIR, with neuroradiologist-examined pixel-wise labels. Here, we use T2-FLAIR for whole tumor segmentation since such modality can better manifest the malignant tumors and is critical to brain surgery of LGG \cite{zeineldin2020deepseg}. In our experiments, the MRI scans are resampled to the same resolution (1 $mm^3$) with intensity normalized to zero mean and unit variance. The data split follows the common settings in the public benchmark \cite{SSL4MIS}, where 250 samples are used for training, 25 for validation and the remaining 60 for testing. 

\subsubsection{Kidney Segmentation Dataset} 
We conduct the experiment on kidney segmentation using the KiTS19 dataset \cite{heller2019kits19}. This dataset collects preoperative 3D abdominal CT images in the late-arterial phase from 210 patients, along with the manual segmentation labels provided by experts. We preprocess the data by means of resampling to 1 $mm^3$ resolution, intensity truncation to $[-75, 175]$ HU followed by intensity normalization, and region-of-interest (ROI) cropping (extracting the 3D patches centering at the kidney). Then, we arbitrarily divide the dataset into three groups: 150 for training, 10 for validation and the remaining 50 for testing. 

\subsubsection{Baseline Approaches}
We compare our method with the supervised-only baselines and several state-of-the-art semi-supervised medical image segmentation methods including: mean-teacher self-ensembling model (MT) \cite{cui2019semi}, uncertainty-aware MT (UA-MT) \cite{yu2019uncertainty}, entropy minimization approach (Entropy Mini) \cite{vu2019EM}, deep adversarial network (DAN) \cite{zhang2017DAN}, interpolation consistency training (ICT) \cite{verma2019ICT} and dual-task mutual learning (DTML) \cite{zhang2021DTML}. We compare those methods using the same backbone and partition protocols to ensure fairness. 

\subsubsection{Implementation Details and Evaluation Metrics}
The framework is implemented in Python with PyTorch, using an NVIDIA GeForce RTX 3090 GPU with 24GB memory. In all experiments, we adopt the same 3D U-Net \cite{3Dunet} as the backbone for a fair comparison. The network is trained using the SGD optimizer (weight decay=$0.0001$, momentum=$0.9$). The batch size is set to 4, including 2 labeled images and 2 unlabeled images in each mini-batch. The final consistency weight $w_{max}$ is empirically set to 0.1, following previous consistency-based methods \cite{yu2019uncertainty}. The maximum number of training steps is set to 20,000 for both tasks. The learning rate is initialized as $0.01$ and decayed with a power of 0.9 after each step. We randomly crop patches of $96 \times 96 \times 96$ voxels as the network input. Standard data augmentation, including randomly cropping, flipping and rotating, is also applied. For a fair comparison, no extra post-processing or ensemble methods are utilized. We use a sliding window strategy with a stride of $64 \times 64 \times 64$ voxels for the inference stage. Then, we adopt four metrics for a comprehensive evaluation, including Dice score, Jaccard, average surface distance (ASD) and 95\% Hausdorff distance (95HD). Two-sided paired t-test with $p \leq 0.05$ is also introduced to test if there is a statistically significant difference in the performance of the proposed method and others.

\begin{table*}[!ht]
\centering
\caption{Quantitative comparison study on the brain tumor segmentation task \cite{brats19}. Standard deviations are shown in parentheses. $*$ indicates $p\leq 0.05$ from a two-sided paired t-test when comparing ours with others. The best mean results are shown in bold.}\label{com_result_brain}
\scalebox{1}{
\begin{tabular}{c|c|c|l|l|l|l}
\Xhline{1pt}
\multirow{2}{*}{Method} & \multicolumn{2}{c|}{\# Training set} & \multicolumn{4}{c}{Metrics}                                                                                                       \\ \cline{2-7} 
                        & Labeled          & Unlabeled         & \multicolumn{1}{c|}{Dice (\%) $\uparrow$} & \multicolumn{1}{c|}{Jaccard (\%) $\uparrow$} & \multicolumn{1}{c|}{95HD (mm) $\downarrow$} & \multicolumn{1}{c}{ASD (mm) $\downarrow$} \\ \hline
Supervised-only                & 100\%             & 0\%                 & 83.84 \scriptsize{(11.93)}                & 74.79 \scriptsize{(15.93)}                   & 8.32 \scriptsize{(9.87)}                         & 2.13 \scriptsize{(1.66)}                   \\\hline                    
Supervised-only                & 10\%             & 0\%                 & 74.43 \scriptsize{(16.67)}$^{*}$                         & 61.86 \scriptsize{(19.62)}$^{*}$                           & 37.11 \scriptsize{(34.12)}$^{*}$                        & 2.79 \scriptsize{(2.08)}$^{*}$                           \\
MT \cite{cui2019semi}                      & 10\%             & 90\%              & 81.94 \scriptsize{(14.53)}$^{*}$                        & 71.67 \scriptsize{(18.51)}$^{*}$                           & 13.62 \scriptsize{(16.05)}$^{*}$                        & 2.33 \scriptsize{(1.99)}$^{*}$                           \\ 
UA-MT \cite{yu2019uncertainty}                   & 10\%             & 90\%              & 80.72 \scriptsize{(16.18)}$^{*}$                        & 70.30 \scriptsize{(19.54)}$^{*}$                           & 11.76 \scriptsize{(13.40)}                & 2.72 \scriptsize{(2.84)}$^{*}$                           \\ 
Entropy Mini \cite{vu2019EM}            & 10\%             & 90\%              & 82.27 \scriptsize{(14.61)}$^{*}$                        & 72.15 \scriptsize{(18.41)}$^{*}$                           & 11.98 \scriptsize{(13.75)}                        & 2.42 \scriptsize{(2.10)}$^{*}$                           \\ 
DAN \cite{zhang2017DAN}                     & 10\%             & 90\%              & 81.71 \scriptsize{(14.92)}$^{*}$                        & 71.43 \scriptsize{(18.79)}$^{*}$                           & 15.15 \scriptsize{(20.38)}$^{*}$                        & 2.32 \scriptsize{(2.15)}$^{*}$                           \\
ICT \cite{verma2019ICT}                     & 10\%             & 90\%              & 77.60 \scriptsize{(22.25)}$^{*}$                        & 67.72 \scriptsize{(24.10)}$^{*}$                           & 15.19 \scriptsize{(18.99)}$^{*}$                        & 3.76 \scriptsize{(4.86)}$^{*}$                           \\ 
DTML \cite{zhang2021DTML}                     & 10\%             & 90\%              & 81.86 \scriptsize{(13.45)}$^{*}$        & 71.83 \scriptsize{(17.58)}$^{*}$           & 16.24 \scriptsize{(21.65)}$^{*}$        & 2.48 \scriptsize{(1.80)}$^{*}$          \\ 
CPCL (ours)             & 10\%             & 90\%              & \textbf{83.36} \scriptsize{(12.58)}                & \textbf{73.23} \scriptsize{(16.43)}                   & \textbf{11.74} \scriptsize{(10.02)}                         & \textbf{1.99} \scriptsize{(1.57)}                   \\ \hline
Supervised-only                & 30\%             & 0\%                 & 78.07 \scriptsize{(11.12)}$^{*}$                        & 66.53 \scriptsize{(14.89)}$^{*}$                           & 28.58 \scriptsize{(11.67)}$^{*}$                        & 2.67 \scriptsize{(1.85)}$^{*}$                           \\ 
MT \cite{cui2019semi}                       & 30\%             & 70\%              & 83.46 \scriptsize{(13.90)}$^{*}$                        & 73.62 \scriptsize{(16.98)}$^{*}$                           & 10.38 \scriptsize{(12.70)}$^{*}$                        & 2.35 \scriptsize{(2.45)}$^{*}$                           \\ 
UA-MT \cite{yu2019uncertainty}                    & 30\%             & 70\%              & 82.63 \scriptsize{(13.29)}$^{*}$                        & 72.28 \scriptsize{(16.69)}$^{*}$                           & 10.01 \scriptsize{(12.13)}$^{*}$                        & 2.42 \scriptsize{(1.97)}$^{*}$                           \\ 
Entropy Mini \cite{vu2019EM}            & 30\%             & 70\%              & 84.75 \scriptsize{(12.85)}$^{*}$                        & 75.31 \scriptsize{(16.07)}                            & 9.10 \scriptsize{(11.44)}                          & 2.16 \scriptsize{(2.23)}$^{*}$                           \\ 
DAN \cite{zhang2017DAN}                     & 30\%             & 70\%              & 84.33 \scriptsize{(13.06)}$^{*}$                        & 74.74 \scriptsize{(16.44)}$^{*}$                           & 10.46 \scriptsize{(16.23)}$^{*}$                        & 2.24 \scriptsize{(2.25)}$^{*}$                           \\ 
ICT \cite{verma2019ICT}                     & 30\%             & 70\%              & 82.17 \scriptsize{(17.01)}$^{*}$                        & 72.49 \scriptsize{(19.45)}$^{*}$                           & \textbf{8.82} \scriptsize{(9.79)}                 & 2.61 \scriptsize{(3.21)}$^{*}$                           \\ 
DTML \cite{zhang2021DTML}                     & 30\%             & 70\%              & 84.81 \scriptsize{(11.17)}$^{*}$        & 75.00 \scriptsize{(14.38)}$^{*}$           & 8.99 \scriptsize{(10.52)}         & 2.02 \scriptsize{(1.72)}           \\ 
CPCL (ours)             & 30\%             & 70\%              & \textbf{85.22} \scriptsize{(11.12)}                         & \textbf{75.68} \scriptsize{(13.89)}                   & 8.97 \scriptsize{(11.65)}                          & \textbf{1.96} \scriptsize{(1.84)}                   \\ \hline
Supervised-only                & 50\%             & 0\%                 & 79.38 \scriptsize{(14.81)}$^{*}$                        & 68.37 \scriptsize{(17.87)}$^{*}$                           & 30.19 \scriptsize{(22.13)}$^{*}$                        & 2.69 \scriptsize{(2.02)}$^{*}$                           \\ 
MT \cite{cui2019semi}                      & 50\%             & 50\%              & 85.67 \scriptsize{(11.58)}$^{*}$                        & 76.46 \scriptsize{(15.18)}$^{*}$                           & \textbf{8.77} \scriptsize{(12.18)}$^{*}$                & 1.93 \scriptsize{(1.82)}                            \\ 
UA-MT \cite{yu2019uncertainty}                    & 50\%             & 50\%              & 84.10 \scriptsize{(14.11)}$^{*}$                        & 74.65 \scriptsize{(17.26)}$^{*}$                           & 8.79 \scriptsize{(11.01)}$^{*}$                         & 2.16 \scriptsize{(2.16)}$^{*}$                           \\ 
Entropy Mini \cite{vu2019EM}            & 50\%             & 50\%              & 85.71 \scriptsize{(11.74)}$^{*}$                        & 76.54 \scriptsize{(15.23)}$^{*}$                           & 9.07 \scriptsize{(11.71)}                          & 1.95 \scriptsize{(1.83)}                            \\ 
DAN \cite{zhang2017DAN}                     & 50\%             & 50\%              & 86.05 \scriptsize{(11.40)}                         & 76.99 \scriptsize{(14.88)}                            & 9.96 \scriptsize{(12.07)}$^{*}$                         & 1.92 \scriptsize{(1.88)}                            \\ 
ICT \cite{verma2019ICT}                     & 50\%             & 50\%              & 82.76 \scriptsize{(15.06)}$^{*}$                        & 72.90 \scriptsize{(18.19)}$^{*}$                           & 9.93 \scriptsize{(11.37)}$^{*}$                         & 2.53 \scriptsize{(2.56)}$^{*}$                           \\ 
DTML \cite{zhang2021DTML}                     & 50\%             & 50\%              & 86.07 \scriptsize{(11.04)}         & 76.67 \scriptsize{(14.60)}$^{*}$          & 9.54 \scriptsize{(11.22)}$^{*}$        & 1.94 \scriptsize{(1.68)}           \\ 
CPCL (ours)             & 50\%             & 50\%              & \textbf{86.41} \scriptsize{(10.08)}                & \textbf{77.29} \scriptsize{(13.79)}                   & 9.04 \scriptsize{(10.88)}                          & \textbf{1.90} \scriptsize{(1.74)}                   \\ \Xhline{1pt}
\end{tabular}}
\end{table*}

\begin{figure*}[!t]
\centerline{\includegraphics[width=2\columnwidth]{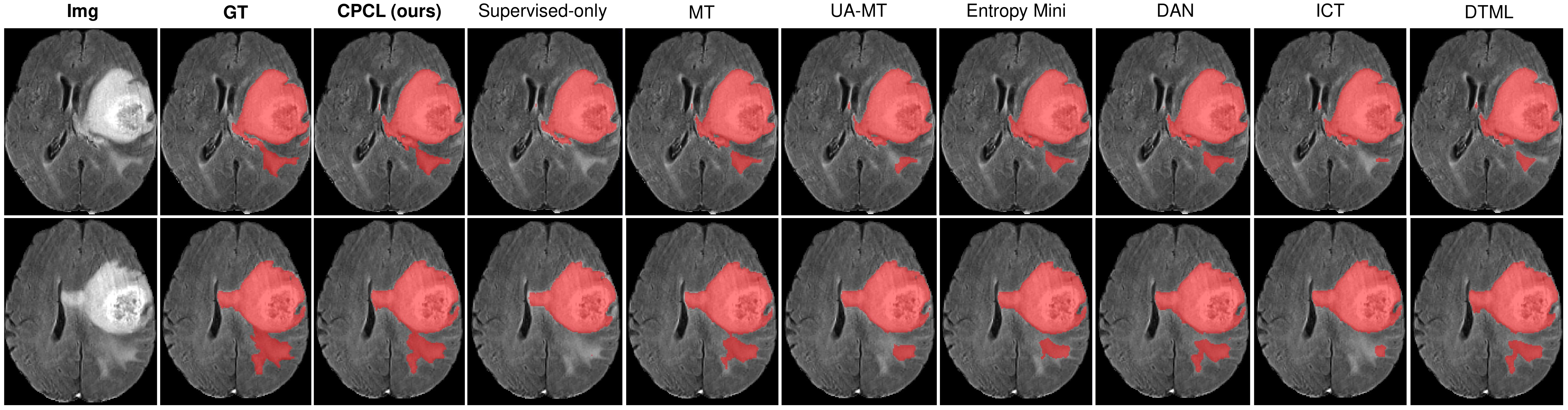}}
\caption{Examples of the whole brain tumor segmentation results of the proposed CPCL and other state-of-the-art approaches under 10\% labeled data setting. Red color represents the segmented whole tumor.}
\label{fig_results_brain}
\end{figure*}

\subsection{Experiments on Brain Tumor Segmentation}
To comprehensively validate the proposed semi-supervised approach, we conduct a comparison study along with rigorous ablation study over the brain tumor segmentation task.

\subsubsection{Comparison Study}
Table \ref{com_result_brain} presents the performance of our method and other state-of-the-art methods under 10\%, 30\% and 50\% labeled data settings. The supervised-only methods serve as the baselines. We observe that segmenting whole brain tumors is much more challenging than the kidney segmentation task (elaborated in Sec. \ref{sec:kidney}) due to the ambiguous tumor boundary and high diversity in tumor appearance. As shown in Table  \ref{com_result_brain},  all semi-supervised methods yield substantial improvements over the supervised-only baselines, revealing that the unlabeled data contains rich and diverse information that can facilitate network learning. Interestingly, we observe that some semi-supervised methods (including ours) can approach or even surpass the supervised-only baseline with 100\% labeled data, implying that the semi-supervised training may provide more productive guidance than the label-only supervision in this challenging task to alleviate the over-fitting issue.

Among the existing methods, MT \cite{cui2019semi}, Entropy Mini \cite{vu2019EM}, DAN \cite{zhang2017DAN} and DTML \cite{zhang2021DTML} achieve more competitive performance. Especially, the Entropy Mini framework trained with only 10\% labeled data achieves the overall largest improvements over the supervised-only baseline in terms of the four metrics. The perturbation-based self-ensembling model, i.e., MT, also shows great performances over the baseline but is slightly worse than Entropy Mini, especially on the 95HD metric. A similar observation is also found on DAN and DTML under 10\% labeled data setting, where their 95HD values are obviously higher than Entropy Mini (15.15 $mm$/16.24 $mm$ vs. 11.98 $mm$). Although the UA-MT obtains slightly worse overall results, it achieves the best 95HD performance over the existing methods under the 10\% labeled data setting. With the increase of labeled data, the performances of all models improve consistently but the gains gradually become limited. Especially, under the most challenging 10\% labeled data setting, the proposed CPCL achieves consistently better performances with most metrics having significant difference (under the two-sided paired t-test) compared with other baselines. Overall, the proposed CPCL outperforms the supervised-only baselines and other state-of-the-art semi-supervised methods with different amounts of labeled data, demonstrating that \textit{``all-around real label supervision"} has a stronger and more reliable capability to exploit the effective information from the unlabeled data. Fig. \ref{fig_results_brain} presents some whole tumor segmentation results of the proposed CPCL and other approaches under 10\% labeled data setting. Consistently, the prediction mask of our proposed CPCL fits more accurately with the ground-truth mask, which further demonstrates the effectiveness of our method.


\begin{table}[t]\scriptsize
\centering
\caption{Ablation study on whole brain tumor segmentation under 10\% labeled data setting. Standard deviations are shown in parentheses. Best results are in bold.}\label{table_ablation_brain}
\scalebox{1}{
\begin{tabular}{c|c|c|c|c}
\Xhline{1pt}
\multirow{2}{*}{Method} & \multicolumn{4}{c}{Metrics}                                                                    \\ \cline{2-5} 
                        & Dice (\%) $\uparrow$ & Jaccard (\%) $\uparrow$ & 95HD (mm) $\downarrow$ & ASD (mm) $\downarrow$ \\ \hline
MT                      & 81.94 \tiny{(14.53)}                         & 71.67 \tiny{(18.51)}                            & 13.62 \tiny{(16.05)}                         & 2.33 \tiny{(1.99)}                  \\ 
F-PCL             & 82.93 \tiny{(13.22)}                & 72.76 \tiny{(17.12)}                   & 12.87 \tiny{(14.69)}                  & 2.03 \tiny{(1.56)}                  \\ 
B-PCL            & 83.16 \tiny{(12.56)}                    & \textbf{73.39} \tiny{(16.58)}                       & 13.20 \tiny{(15.95)}                      & 2.00 \tiny{(1.55)}                     \\ 
MT-F-PCL          & 82.72 \tiny{(12.96)}                    & 72.40 \tiny{(17.06)}                       & 14.03 \tiny{(17.60)}                      & 2.11 \tiny{(1.64)}                     \\ 
MT-B-PCL         & 82.90 \tiny{(13.09)}                    & 72.70 \tiny{(17.19)}                       & 15.32 \tiny{(16.57)}                      & 2.09 \tiny{(1.70)}                     \\ 
MT-C-PCL         & 82.86 \tiny{(13.68)}                &    72.79 \tiny{(17.64)}                   & 13.34 \tiny{(17.28)}                     & 2.14 \tiny{(1.78)}                     \\ 
CPCL                   & \textbf{83.36} \tiny{(12.58)}                & 73.23 \tiny{(16.43)}                   & \textbf{11.74} \tiny{(10.02)}                         & \textbf{1.99} \tiny{(1.57)}         \\ \Xhline{1pt}
\end{tabular}}
\end{table}

\subsubsection{Analytical Ablation Study}
\label{sec:ablation}
Our CPCL framework partly benefits from the self-ensembling strategy. Besides the typical self-ensembling mean-teacher (MT) model, we further propose different variants to perform an ablation study under the most challenging 10\% labeled data setting: a) \textbf{F-PCL}: only preserving the forward prototype consistency learning, i.e., $\mathcal{L}=\mathcal{L}_{s}+\lambda \mathcal{L}_{fpc}$; b) \textbf{B-PCL}: only preserving the backward prototype consistency, i.e., $\mathcal{L}=\mathcal{L}_{s}+\lambda \beta\mathcal{L}_{bpc}$; c) \textbf{MT-F-PCL}: combining the previous perturbation-based consistency (adding random Gaussian noises) and our forward consistency; d) \textbf{MT-B-PCL}: combining the previous perturbation-based consistency and our backward consistency; e) \textbf{MT-C-PCL}: combining the previous perturbation-based consistency and our cyclic prototype consistency.

As shown in Table \ref{table_ablation_brain}, both the forward and backward prototype learning mechanisms can independently contribute to the performance gains compared to the previous perturbation-based consistency method (i.e., the MT model). Specifically, the improvement brought by F-PCL demonstrates that the real label prototypes bring more effective and meaningful knowledge for training the segmentation network than that of previous perturbed unsupervised targets. Interestingly, it can be observed that B-PCL achieves more competitive results compared to F-PCL. Let us refer back to Fig. \ref{fig_framework}, we can find that the prototypical prediction of the backward process relies on well-rounded $P_{u}$, $F_{u}$ and $F_{l}$, and has a more direct interaction with the real label $Y_{l}$, while $Y_{l}$ in the forward process serves as an oblique but critical role for prototype extraction. Such an observation indicates that more direct real label supervision can help learn more effective information in our framework. Besides, we can observe that combining our prototype consistency with the previous perturbation-based consistency (MT-F-PCL, MT-B-PCL and MT-C-PCL) can also improve the overall performance but not as much as only using our prototype consistency mechanism (F-PCL, B-PCL and CPCL). Particularly, MT-F-PCL, MT-B-PCL and MT-C-PCL obtain relatively worse 95HD than only using our prototype consistency. Therefore, we recommend not mixing the two mechanisms. Thanks to the complementary forward and backward processes, the overall superior and robust performance can be achieved via our CPCL framework. The ablation study further demonstrates that CPCL can serve as a superior alternative to the previous perturbation-based consistency method.

\subsubsection{Impact of Different Loss Weight $\beta$}
\label{sec:beta}

\begin{table}[t]\scriptsize
\centering
\caption{Results of whole brain tumor segmentation with different loss weight $\beta$ under 10\% labeled data setting. Standard deviations are shown in parentheses. Best results are in bold.}\label{table_beta_brain}
\scalebox{1}{
\begin{tabular}{c|c|c|c|c}
\Xhline{1pt}
\multirow{2}{*}{$\beta$} & \multicolumn{4}{c}{Metrics}                                                                    \\ \cline{2-5} 
                        & Dice (\%) $\uparrow$ & Jaccard (\%) $\uparrow$ & 95HD (mm) $\downarrow$ & ASD (mm) $\downarrow$ \\ \hline
1                      & 82.89 \tiny{(13.10)}                & 72.67 \tiny{(17.07)}                   & 15.32 \tiny{(22.14)}                  & 2.01 \tiny{(1.67)}                  \\ 
5             & 83.24 \tiny{(13.05)}                & 73.18 \tiny{(17.01)}                   & \textbf{11.47} \tiny{(14.21)}                  & 2.02 \tiny{(1.60)}                  \\ 
10            & \textbf{83.36} \tiny{(12.58)}                & \textbf{73.23} \tiny{(16.43)}                   & 11.74 \tiny{(10.02)}                         & \textbf{1.99} \tiny{(1.57)}                     \\ 
15          & 83.29 \tiny{(12.58)}                   & 73.14 \tiny{(16.54)}                      & 14.10 \tiny{(18.60)}                     & 2.00 \tiny{(1.60)}                    \\ 
20         & 82.87 \tiny{(13.39)}                   & 72.78 \tiny{(17.47)}                      & 13.21 \tiny{(18.48)}                     & 2.11 \tiny{(1.74)}                    \\ \Xhline{1pt}
\end{tabular}}
\end{table}

The ablation study indicates that the semi-supervised training benefits more from the backward process than the forward process. As shown in our loss function (Eqn. (\ref{eq:total_loss})), a fixed factor $\beta$ is introduced to control the trade-off between the forward process and backward process. Here, we investigate the impact of different $\beta$, and the results are shown in Table \ref{table_beta_brain}. It can be observed that the proposed CPCL is not particularly sensitive to $\beta$, except for the 95HD metric, but overall, it performs optimally when $\beta = 10$. Thus, we set $\beta = 10$ in all experiments involving the prototypical backward process.

\begin{table*}[!ht]
\centering
\caption{Quantitative comparison study on the kidney segmentation task \cite{heller2019kits19}. Standard deviations are shown in parentheses. $*$ indicates $p\leq 0.05$ from a two-sided paired t-test when comparing ours with others. The best mean results are in bold.}\label{com_result_kits}
\scalebox{1}{
\begin{tabular}{c|c|c|l|l|l|l}
\Xhline{1pt}
\multirow{2}{*}{Method} & \multicolumn{2}{c|}{\# Training set} & \multicolumn{4}{c}{Metrics}                                                                                                       \\ \cline{2-7} 
                        & Labeled          & Unlabeled         & \multicolumn{1}{c|}{Dice (\%) $\uparrow$} & \multicolumn{1}{c|}{Jaccard (\%) $\uparrow$} & \multicolumn{1}{c|}{95HD (mm) $\downarrow$} & \multicolumn{1}{c}{ASD (mm) $\downarrow$} \\ \hline
Supervised-only                & 100\%             & 0\%                 & 96.51 \scriptsize{(5.02)}                        & 93.64 \scriptsize{(8.77)}                           & 3.17 \scriptsize{(3.09)}                        & 0.41 \scriptsize{(0.40)}                          \\\hline
Supervised-only                & 5\%             & 0\%                 & 89.64 \scriptsize{(8.66)}$^{*}$                        & 82.40 \scriptsize{(12.71)}$^{*}$                           & 10.64 \scriptsize{(8.61)}$^{*}$                        & 0.79 \scriptsize{(0.61)}$^{*}$                          \\
MT \cite{cui2019semi}                      & 5\%             & 95\%              & 92.92 \scriptsize{(7.23)}                         & 87.78 \scriptsize{(10.96)}$^{*}$                           & 6.13 \scriptsize{(7.18)}$^{*}$                        & 0.64 \scriptsize{(0.48)}                            \\ 
UA-MT \cite{yu2019uncertainty}                   & 5\%             & 95\%              & 92.88 \scriptsize{(6.89)}$^{*}$                        & 87.63 \scriptsize{(10.58)}$^{*}$                           & 6.57 \scriptsize{(7.14)}$^{*}$               & 0.62 \scriptsize{(0.46)}                            \\ 
Entropy Mini \cite{vu2019EM}            & 5\%             & 95\%              & 92.76 \scriptsize{(7.01)}$^{*}$                       & 87.01 \scriptsize{(10.65)}$^{*}$                           & 6.45 \scriptsize{(6.99)}$^{*}$                        & 0.69 \scriptsize{(0.48)}$^{*}$                           \\ 
DAN \cite{zhang2017DAN}                     & 5\%             & 95\%              & 92.87 \scriptsize{(6.89)}$^{*}$                       & 87.65 \scriptsize{(10.56)}$^{*}$                          & 6.39 \scriptsize{(7.25)}$^{*}$                       & 0.68 \scriptsize{(0.45)}$^{*}$                          \\ 
ICT \cite{verma2019ICT}                     & 5\%             & 95\%              & 92.47 \scriptsize{(6.88)}$^{*}$                       & 86.97 \scriptsize{(10.86)}$^{*}$                          & 7.20 \scriptsize{(8.42)}$^{*}$                       & 0.73 \scriptsize{(0.62)}$^{*}$                          \\ 
DTML \cite{zhang2021DTML}                     & 5\%             & 95\%              & 92.54 \scriptsize{(9.50)}$^{*}$        & 87.27 \scriptsize{(13.07)}$^{*}$          & 6.90 \scriptsize{(8.53)}$^{*}$       & 0.67 \scriptsize{(0.61)}$^{*}$         \\ 
CPCL (ours)             & 5\%             & 95\%              & \textbf{93.43} \scriptsize{(7.19)}                & \textbf{88.67} \scriptsize{(10.51)}                   & \textbf{5.33} \scriptsize{(6.52)}                         & \textbf{0.59} \scriptsize{(0.56)}                   \\ \hline
Supervised-only                & 10\%             & 0\%                 & 92.31 \scriptsize{(7.60)}$^{*}$                        & 86.72 \scriptsize{(11.22)}$^{*}$                           & 6.84 \scriptsize{(7.48)}$^{*}$                        & 0.67 \scriptsize{(0.55)}$^{*}$                           \\ 
MT \cite{cui2019semi}                       & 10\%             & 90\%              & 93.98 \scriptsize{(6.78)}$^{*}$                        & 89.81 \scriptsize{(10.09)}$^{*}$                           & 4.63 \scriptsize{(6.28)}$^{*}$                        & 0.56 \scriptsize{(0.45)}                            \\ 
UA-MT \cite{yu2019uncertainty}                    & 10\%             & 90\%              & 94.12 \scriptsize{(6.89)}                         & 90.02 \scriptsize{(10.16)}$^{*}$                           & 4.52 \scriptsize{(6.36)}$^{*}$                        & 0.56 \scriptsize{(0.53)}                            \\ 
Entropy Mini \cite{vu2019EM}            & 10\%             & 90\%              & 94.05 \scriptsize{(6.06)}                         & 90.36 \scriptsize{(9.28)}                            & 4.34 \scriptsize{(6.24)}                          & 0.55 \scriptsize{(0.43)}                            \\ 
DAN \cite{zhang2017DAN}                     & 10\%             & 90\%              & 93.94 \scriptsize{(6.86)}$^{*}$                       & 89.65 \scriptsize{(10.27)}$^{*}$                          & 4.79 \scriptsize{(6.31)}$^{*}$                       & 0.59 \scriptsize{(0.50)}$^{*}$                          \\ 
ICT \cite{verma2019ICT}                     & 10\%             & 90\%              & 94.02 \scriptsize{(4.98)}                         & 89.58 \scriptsize{(8.23)}$^{*}$                          & 4.40 \scriptsize{(6.43)}                & 0.61 \scriptsize{(0.57)}                           \\ 
DTML \cite{zhang2021DTML}                     & 10\%             & 90\%              & 94.04 \scriptsize{(9.67)}$^{*}$       & 89.88 \scriptsize{(12.62)}$^{*}$          & 4.86 \scriptsize{(7.28)}$^{*}$       & 0.57 \scriptsize{(0.55)}          \\ 
CPCL (ours)             & 10\%             & 90\%              & \textbf{94.59} \scriptsize{(5.96)}                         & \textbf{90.69} \scriptsize{(9.19)}                   & \textbf{4.15} \scriptsize{(5.94)}                          & \textbf{0.54} \scriptsize{(0.43)}                   \\ \Xhline{1pt}
\end{tabular}}
\end{table*}

\subsection{Experiments on Kidney Segmentation}
\label{sec:kidney}
To further validate the proposed approach, we also conduct experiments on kidney segmentation from abdominal CT images. All the implementation settings are consistent with the above whole brain tumor segmentation task, except for the partition protocols. 

\begin{figure}[t]
\centerline{\includegraphics[width=\columnwidth]{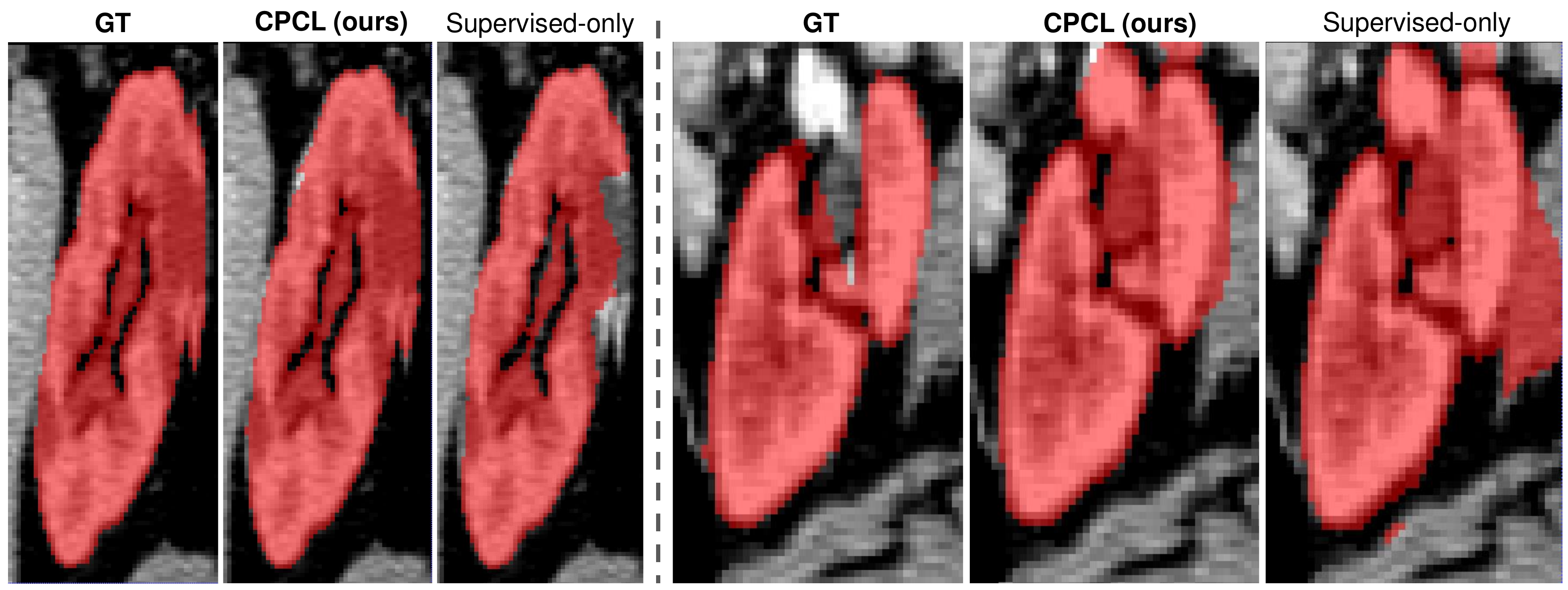}}
\caption{Examples of the kidney segmentation results of the proposed CPCL and the supervised-only baseline under 5\% labeled data setting. Red color represents the segmented kidney.}
\label{fig_results_kidney}
\end{figure}

Table \ref{com_result_kits} presents the performance of our method and other semi-supervised methods under 5\% and 10\% labeled data settings, respectively. Such training set configuration depends on our empiricism that segmenting kidneys is much easier than the whole brain tumor segmentation. As shown in Table \ref{com_result_kits}, with only 5\% labeled data, the supervised-only method yields an average Dice score of 89.64\%, which is difficult to achieve in the former whole brain tumor segmentation task. Although the performances are very comparable among all the methods due to less difficulty of the segmentation targets, the proposed CPCL can still achieve the overall best results in terms of four metrics under the same partition protocol, which further demonstrates the superiority and robustness of our approach. Especially, under the least labeled data setting, i.e., 5\%, the proposed CPCL further advances performances with most metrics having significant difference (under the two-sided paired t-test) compared with other baselines. Fig. \ref{fig_results_kidney} presents two exemplary kidney segmentation results of the proposed CPCL and the supervised-only baseline under the 5\% labeled data setting. We can see that the proposed CPCL achieves visually better segmentation results compared to the supervised-only baseline.

\section{Discussions}
\begin{figure}[t]
\centerline{\includegraphics[width=\columnwidth]{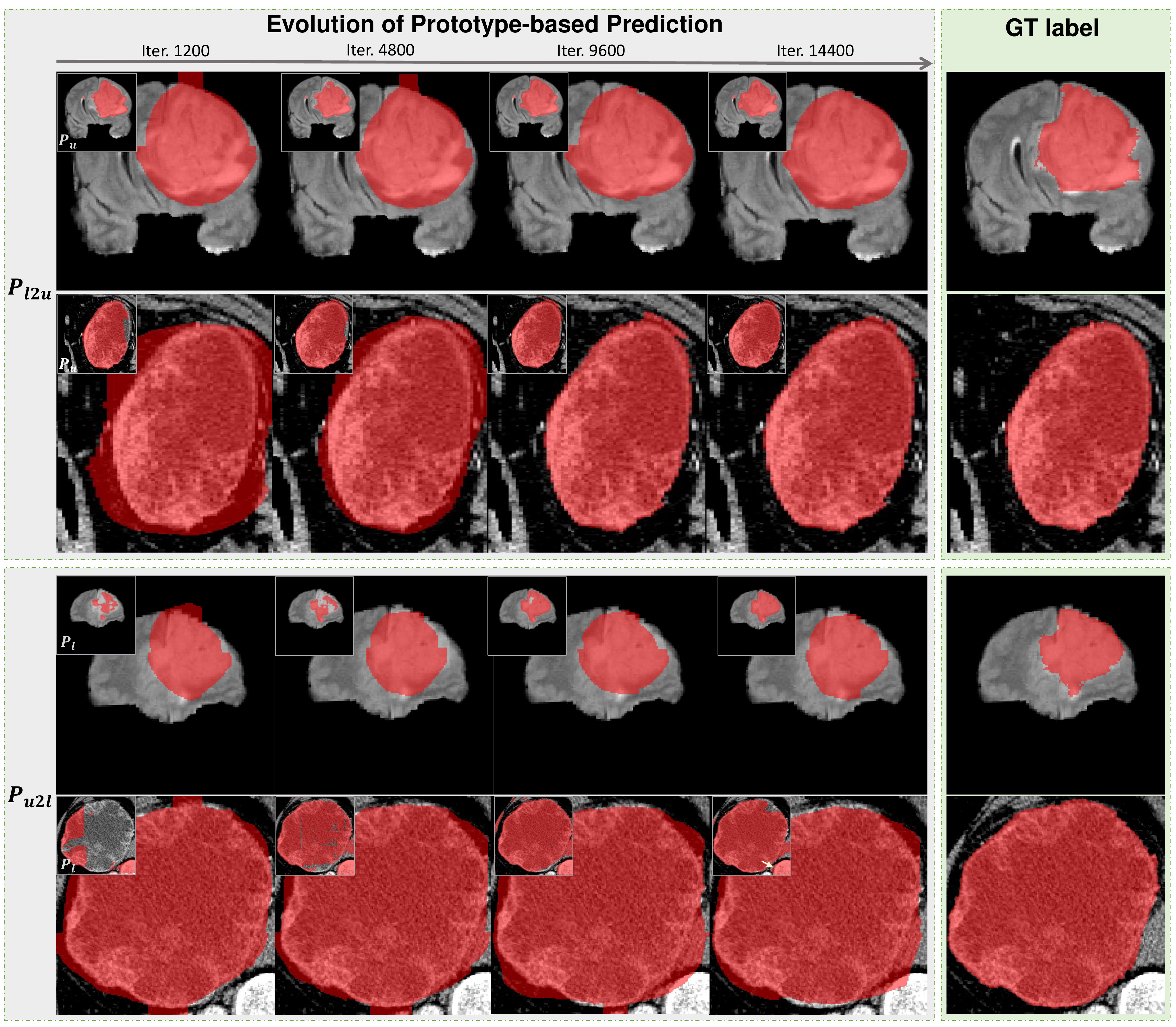}}
\caption{Exemplar evolution of the prototype-based predictions $P_{l2u}$ and $P_{u2l}$ (with model predictions $P_u$/$P_l$ shown on the top-left corner) under 10\% (whole brain tumor segmentation) and 5\% (kidney segmentation) labeled data settings during the training process. Red color represents the segmented targets.}
\label{fig_evolution}
\end{figure}

To better understand the learning behavior and visually evaluate the quality of the prototype in our proposed CPCL, we visualize the evolution of the prototype-based predictions under 10\% (whole brain tumor segmentation) and 5\% (kidney segmentation) labeled data settings in different training periods, as shown in Fig. \ref{fig_evolution}. The corresponding model predictions $P_{u}$ and $P_{l}$ are also visualized at the top-left corner with the prototype-based predictions $P_{l2u}$ and $P_{u2l}$, respectively. At the early training stage, it is observed that the model predictions tend to under-segment the objects, yet the prototype-based predictions often indicate more accurate target regions. Overall, as the training goes on, both the model predictions and prototypical predictions are gradually refined, indicating that the learned features become more discriminative and compact. Ideally, we hope the prototypical prediction at the late training stage can be perfectly consistent with the ground-truth segmentation. However, since the holistic prototype mainly relies on an average prior via the masked average pooling approach, therefore the prototypical predictions lean to predict a relatively large and coarse area for the segmentation target. Interestingly, due to the ambiguous tumor boundary and variegated intensity within the tumor area, the above observation is particularly evident in the whole brain tumor segmentation task. Despite the imperfectness, the network training can still benefit from the real label-centric supervision mechanism and acquire effective information from those real label-driven consistency targets, as experimentally demonstrated above. These findings also echo the previous works \cite{ning2020macro} that the target-related weak labels can offer more high-level region proposals for segmentation. However, intuitively, improving the quality of prototypical prediction can better impose real label supervision.  Besides, the ablation study also indicates that our backward process has more direct interaction with the real labels. Thus, for future work, we may consider removing the forward process to construct a more elegant framework and pay more attention to improving the discriminability and compactness of the class prototypes. Since the spirit in few-shot segmentation (FSS) is in line with the proposed perspective in semi-supervised segmentation, i.e., learning transferable knowledge from support set (labeled set) to query set (unlabeled set), more collaboration between the two research fields will be interesting future work on top of our early attempt. For example, introducing the superpixel technique to form an adaptive prototype allocation mechanism \cite{li2021adaptive} or considering the feature differences between foreground and background \cite{nguyen2019FWB} can be effective ways to further improve the prototype quality and thus enhance the final performance.

Besides the limitation caused by the holistic prototype, another challenge is that we assume both labeled and unlabeled samples are from the same distribution. However, the abundant image-only data are often collected from different devices and clinical centers, where the resulting domain shift will cause substantial performance degradation in semi-supervised learning \cite{oliver2018realistic}. Such a limitation also applies to most of current semi-supervised methods. Particularly, the domain shift may harm the extracted features in our framework, which may disastrously mislead the prototype extraction and thereby interfere with the cyclic prototype consistency learning process. Therefore, further investigating how to adapt domain adaptation to deal with potential domain shifts in our framework is also an interesting direction with fruitful clinical values.

Overall, this work reveals that our initial attempt, i.e., utilizing expert-examined real labels to explicitly supervise the network learning from both their paired labeled data and unpaired unlabeled data, is challenging yet feasible. Compared with the previous unsupervised consistency fashion, radiologists may have more confidence in the trained model of our framework since we explicitly leverage the real labels acknowledged by the experts themselves during the whole network training. We hope that this work can evolve into a new direction in semi-supervised segmentation and inspire more future research devoted to more reliable models and making more effective use of the precious high-quality labeled data.

\section{Conclusion}
In this work, we studied the semi-supervised segmentation task from an unexplored perspective, i.e., exploiting the unlabeled data via explicit real label supervision. To this end, we proposed a novel cyclic prototype consistency learning (CPCL) framework constructed by a labeled-to-unlabeled prototypical forward process and an unlabeled-to-labeled backward process. In this way, our framework turns exisiting \textit{``unsupervised"} consistency into new \textit{``supervised"} consistency, obtaining the \textit{``all-around real label supervision"} property of our method. Extensive experiments on two public datasets demonstrated the superiority of our method over other state-of-the-art semi-supervised learning methods on both whole brain tumor segmentation from T2-FLAIR MRI and kidney segmentation from CT images.

\bibliographystyle{IEEEtran.bst}
\bibliography{IEEEexample.bib} 

\end{document}